\documentclass[%
 aip,
 amsmath,amssymb,
 reprint,%
]{revtex4-1}

\usepackage{graphicx}% Include figure files
\usepackage{dcolumn}% Align table columns on decimal point
\usepackage{bm}% bold math
\usepackage[utf8]{inputenc}
\usepackage[T1]{fontenc}
\usepackage{mathptmx}
\usepackage{etoolbox}

\makeatletter
\def\@email#1#2{%
 \endgroup
 \patchcmd{\titleblock@produce}
  {\frontmatter@RRAPformat}
  {\frontmatter@RRAPformat{\produce@RRAP{*#1\href{mailto:#2}{#2}}}\frontmatter@RRAPformat}
  {}{}
}%
\makeatother
\begin{document}

\preprint{AIP/123-QED}

\title{Comparison and analysis of methods for measuring the spin transverse relaxation time of rubidium atomic vapor}
% Force line breaks with \\
\author{Lulu Zhang}

\author{Ni Zhao}
\author{Yongbiao Yang}
\author{Junye Zhao}
\affiliation{ 
State Key Laboratory of Quantum Optics and Quantum Optics Decices, Shanxi University, Institute of Opto-Electronics,Taiyuan 030006, China %\\This line break forced with \textbackslash\textbackslash
}%

\author{Jun He}
\author{Junmin Wang}
 \homepage{Corresponding author: wwjjmm@sxu.edu.cn; ORCID : 0000-0001-8055-000X}
\affiliation{ 
State Key Laboratory of Quantum Optics and Quantum Optics Decices, Shanxi University, Institute of Opto-Electronics,Taiyuan 030006, China %\\This line break forced with \textbackslash\textbackslash
}%

\affiliation{ 
Collaborative Innovation Center of Extreme Optics, Shanxi University, Taiyuan 030006, China%\\This line break forced with \textbackslash\textbackslash
}%

\begin{abstract}
The spin transverse relaxation time (${T_{2}}$) of atoms is an important indicator for precision measurement. Several methods have been proposed to characterize the \textbf{${T_{2}}$} of atoms. In this paper, the ${T_{2}}$ of rubidium (Rb) atomic vapor in the same cell was measured using four measuring methods, namely spin noise spectrum signal fitting, improved free induction decay (FID) signal fitting, \textbf{$\omega_m$}-broadening fitting, and magnetic resonance broadening fitting. Meanwhile, the ${T_{2}}$ of five different types of Rb atomic vapor cells were measured and characterized. A comparative analysis visualizes the characteristics of the different measuring methods and the effects of buffer gas on ${T_{2}}$ of Rb. We theoretically and experimentally analyzed the applicability of the different methods, and then demonstrated that the improved FID signal fitting method provides the most accurate measurement because of the clean environment in which the measurements were taken. Furthermore, we demonstrated and qualitatively analyzed the relationship between the atomic number density and the ${T_{2}}$ of Rb. This work provides analytical insight in selecting atomic vapor cells, and may shed light on the improvement of the sensitivity of atomic magnetometers. 
\end{abstract}

\maketitle

\section{\label{sec:level1}Introduction}

The  precise measurement of magnetic fields has very important application potential in basic physics research [1], [2], geophysics [3], clinical medicine [4], and cosmic dark matter measurement [5]. The transverse relaxation time (\textbf{${T_{2}}$}) of atoms is an important indicator to characterize the performance of magnetic field precision measurements. Especially, in atomic magnetometers [6],[7],[8], the sensitivity improves with \textbf{${T_{2}}$} of atoms. Because the \textbf{${T_{2}}$} of atoms contains a large amount of information, it is necessary to accurately measure the \textbf{${T_{2}}$} of atoms to help select the ideal type of atomic vapor cell.

Currently, the traditional free induction decay(FID) method is a commonly used method for measuring the \textbf{${T_{2}}$} of atoms. In 2017, Jiang et al [9] proposed the fitting-ratio method and the magnetic resonance broadening method to measure the \textbf{${T_{2}}$} of atoms, and compared with the traditional FID method to explore the advantages of the new methods. In 2020, the group proposed a perturbation-free method to measure the \textbf{${T_{2}}$} [10] of atoms in nuclear magnetic resonance oscillator. In 2023, Wei et al [11] measured the \textbf{${T_{2}}$} of cesium using three methods.

In addition, the spin noise spectrum signal in the spin noise spectroscopy technique also contains a significant information, and is an important method for the measurement of the \textbf{${T_{2}}$}. In 2007, Katsoprinakis et al [12] analyzed the spin noise spectrum properties of atoms in detail from theoretical and experimental perspectives, revealing the relationship between the spin spectrum and the \textbf{${T_{2}}$} of atoms.

In this paper, we select spin noise spectrum fitting, improved FID signal fitting, \textbf{$\omega_m$}-broadening fitting and magnetic resonance broadening fitting, which are the four methods for measuring the \textbf{${T_{2}}$} of atoms. In addition, five types of rubidium (Rb) vapor cells are selected to measure their \textbf{${T_{2}}$}, to visually observe the merits of various methods and the effects of different types of buffer gases on the \textbf{${T_{2}}$} of Rb through the comparative analysis of the experimental results.

\section{Theoretical analysis}
The relaxation time of an atom is divided into longitudinal relaxation time (\textbf{${T_{1}}$}) and transverse relaxation time. Longitudinal relaxation is the relaxation of the atomic state population to a certain equilibrium value, which is related to the spin of the atom, that is, the lifetime of the atomic state. Transverse relaxation is phase-dependent and refers to the decoherence time of phase.

For alkali metal atomic ensembles, under the action of the static magnetic field ${B_{z}}$ along the z-axis, the components of macroscopic magnetization of atoms in thermal equilibrium are:

\begin{equation}\label{eq1}
\begin{split}
\textbf{${M_{z}}$}=\textbf{${M_{0}}$},\\
\textbf{${M_{x}}$}=\textbf{${M_{y}}$}=0.
\end{split}
\end{equation}

Here ${M_0}$ is a constant value. If a pump light is simultaneously applied along the z-axis to polarize the alkali metal atoms, an oscillating magnetic field with amplitude ${B_{1}}$, frequency \textbf{$\Omega$} is introduced, whose direction is perpendicular to the static magnetic field ${B_{z}}$, to deflect the spin polarization of the alkali metal atoms away from the z-axis. Then the pump light and the oscillating magnetic field are withdrawn, the alkali metal atoms are in a non-equilibrium state, and some relaxation mechanism restores it to a thermal equilibrium state. In a rotating coordinate system, the evolution of the macroscopic magnetization of alkali metal atoms can be expressed by the Bloch equation, which is described as follows [13]: 

\begin{equation}\label{eq}
\begin{split}
\frac{d\textbf{$M'_{x}$}(t)}{dt}=\Delta\omega\textbf{$M'_{y}$}(t)-\frac{\textbf{$M'_{x}$}(t)}{\textbf{${T_{2}}$}},\\
\frac{d\textbf{$M'_{y}$}(t)}{dt}=\gamma\textbf{${B_{1}}$}\textbf{$M_{z}$}{(t)}-\Delta\omega\textbf{$M'_{x}$}(t)-\frac{d\textbf{$M'_{y}$}(t)}{\textbf{${T_{2}}$}},\\
\frac{\textbf{$M_{z}$}(t)}{dt}=-\gamma\textbf{${B_{1}}$}\textbf{$M'_{y}$}{(t)}+\frac{\textbf{$M_{0}$}-\textbf{$M_{z}$}(t)}{\textbf{${T_{1}}$}}.
\end{split}
\end{equation}

Here, ${\Delta\omega = \omega_{L} - \Omega}$ is the mismatch between the Larmor frequency ${\omega_{L}}$ and the oscillating magnetic field frequency ${\Omega}$, and $\gamma$ is the ground-state gyromagnetic ratio. In this process, the macroscopic magnetization of alkali metal atoms is decomposed into a component $M_{z}$ (parallel to the static magnetic field) and components $M_{x}$, $M_{y}$ (perpendicular to the static magnetic field). The characteristic time from $M_{z}$ to $M_{0}$ is referred to as the ${T_{1}}$. The characteristic time for $M_{x}$ and $M_{y}$ to gradually return to 0 is called the ${T_{2}}$, in which the spin precession phase of the alkali metal atoms is redistributed until it is disordered.

\subsection{Spin noise spectrum fitting}

Spin noise is the random distribution of atomic electron spins in quasi-thermodynamic equilibrium. The spin noise spectrum is an optical technique that can be obtained from nuclear magnetic resonance measurements and magnetic force microscopy measurements; however, the most sensitive and widely used detection technique is Faraday rotation, which maps atomic spin noise on the polarization plane of a non-resonant probe light. Furthermore, we experimentally measured and analyzed the spin noise spectrum of Rb in previous articles [14], [15], and have a certain understanding of its principle and parameter optimization.

Based on the principle of the spin noise spectrum, the spin noise spectrum signal of Voigt configuration correlate with ${T_2}$. The light beam is perpendicular to the static magnetic field, the random fluctuation of magnetization process around the direction of the static magnetic field, and this fluctuation on the spectrum manifests as Lorentzian-linear peak. The full width at half maximum (FWHM) of the peak is inversely correlated with the ${T_2}$ [16],[17]. Moreover, the experimental setup used in our apparatus employed this method to measure the ${T_2}$.

\subsection{Improved FID signal fitting}

In this method, an all-optical Bell-Bloom magnetometer [18] modulated with pump light and a magnetometer driven by radio frequency (RF) magnetic field [19] are the two commonly used experiment setups. In Bell-Bloom magnetometer, the direction of the static magnetic field is perpendicular to the pump light, which is modulated in frequency or amplitude. In RF magnetometer, the direction of the pump light is parallel to the static magnetic field, and an oscillating magnetic field with the Larmor frequency is used for \textbf{$\pi$}/2 pulse times. These two approaches cause the macroscopic magnetization of the atomic ensemble in a plane perpendicular to the static magnetic field, and the precession is exponentially attenuated around the static magnetic field at the Larmor frequency. This Lamor precession is mapped to the rotation of the probe light polarization plane, and the FID signal is subsequently detected with a differential detector. The \textbf{${T_{2}}$} of atoms can be obtained [20], [21] simply by fitting the envelope of the FID signal:

\begin{equation}\label{eq}
\begin{split}
\textbf{${S_{FID}}$}=\textbf{${A_{FID}}$}exp(-t/\textbf{${T_{2}}$})
\end{split}
\end{equation}

Here, we improve and optimize the traditional method of obtaining the FID signal based on the RF magnetometer, that is, the action time of pump light, probe light, and RF magnetic field in the time domain are controlled by timing sequence, to avoid the signal being affected by the pump light and the additional magnetic field during the detection [22].

\subsection{\textbf{$\omega_m$}-broadening fitting}

In this method, a modulated magnetic field with frequency $\omega_m$ is applied in the direction of the static magnetic field \textbf{${B_{z}}$}. If there are magnetic fields \textbf{${B_{x}}$} and \textbf{${B_{y}}$} considerable smaller than static magnetic field \textbf{${B_{z}}$} exists in the x and y directions, the expression of the in-phase and quadrature phase signals after lock-in amplifier(LIA) demodulation is [9]:

\begin{equation}\label{eq}
\begin{split}
\textbf{${S_{IS}}$}\propto\frac{\textbf{${B_{y}}$}-\textbf{${B_{x}}$}(\gamma\textbf{${B_{z}}$}+n\textbf{$\omega_m$})\textbf{${T_{2}}$}}{1+\textbf{${T_{2}}$}^2\left(\gamma\textbf{${B_{z}}$}+n\textbf{$\omega_m$}\right)^2},\\
\textbf{${S_{QS}}$}\propto\frac{\textbf{${B_{x}}$}-\textbf{${B_{y}}$}(\gamma\textbf{${B_{z}}$}+n\textbf{$\omega_m$})\textbf{${T_{2}}$}}{1+\textbf{${T_{2}}$}^2\left(\gamma\textbf{${B_{z}}$}+n\textbf{$\omega_m$}\right)^2}.
\end{split}
\end{equation}

where n is the ratio of the Larmor frequency \textbf{$\omega_L$} to the modulation frequency \textbf{$\omega_m$}, i.e., n= \textbf{$\omega_0$} / \textbf{$\omega_m$}. When \textbf{${B_{x} \neq 0}$}, \textbf{${B_{y} = 0}$}, the above equation can be written as:

\begin{equation}\label{eq}
\begin{split}
\textbf{${S_{IS}}$}\propto\frac{\textbf{${B_{x}}$}(\gamma\textbf{${B_{z}}$}+n\textbf{$\omega_m$})\textbf{${T_{2}}$}}{1+\textbf{${T_{2}}$}^2\left(\gamma\textbf{${B_{z}}$}+n\textbf{$\omega_m$}\right)^2},\\
\textbf{${S_{QS}}$}\propto\frac{\textbf{${B_{x}}$}}{1+\textbf{${T_{2}}$}^2\left(\gamma\textbf{${B_{z}}$}+n\textbf{$\omega_m$}\right)^2}.
\end{split}
\end{equation}

Similarly, we can find the in-phase and quadrature phase signals of the LIA demodulation in case \textbf{${B_{x} = 0}$}, \textbf{${B_{y} \neq 0}$}. From the above equation, the relationship between the FWHM ($\Gamma$) and the ${T_{2}}$ can be obtained: \textbf{$\Gamma_{QS}$} = \textbf{$\Gamma_{IS}$} = 2/ (n${T_{2}}$).

\subsection{magnetic resonance broadening fitting}

In this method, an RF magnetic field with magnitude \textbf{${B_{1}}$}, frequency \textbf{$\Omega$} is applied perpendicular to the direction of the static magnetic field. In a rotating coordinate system, the steady-state solution of the magnetization \textbf{$M$} is [23]:

\begin{equation}\label{eq}
\begin{split}
\textbf{$M_{x}$}= \frac{\textbf{$M_{0}$}\gamma\textbf{$B_{1}$}\Delta\omega}{(1/\textbf{${T_{2}}$})^2+(\Delta\omega)^2+(\textbf{${T_{1}}$}/\textbf{${T_{2}}$})(\gamma\textbf{${B_{1}}$})^2},\\
\textbf{$M_{y}$}= \frac{\textbf{$M_{0}$}\gamma\textbf{$B_{1}$}(\textbf{${1/T_{2}}$})}{(1/\textbf{${T_{2}}$})^2+(\Delta\omega)^2+(\textbf{${T_{1}}$}/\textbf{${T_{2}}$})(\gamma\textbf{${B_{1}}$})^2},\\
\textbf{$M_{z}$}= \frac{\textbf{$M_{0}$}[(\Delta\omega)^2+(1/\textbf{${T_{2}}$})^2]}{(1/\textbf{${T_{2}}$})^2+(\Delta\omega)^2+(\textbf{${T_{1}}$}/\textbf{${T_{2}}$})(\gamma\textbf{${B_{1}}$})^2}.
\end{split}
\end{equation}

 Scanning the static magnetic field \textbf{${B_{0}}$}, the relationship between the FWHM and the ${B_{1}}$ after LIA demodulation is [9] :

\begin{equation}\label{eq}
\begin{split}
\textbf{$\Gamma$} = 2\sqrt{(1/\textbf{${T_{2}}$})^2+\textbf{${T_{1}}$}/\textbf{${T_{2}}$})(\gamma\textbf{${B_{1}}$})^2}
\end{split}
\end{equation}

And \textbf{$\Gamma = 2/ T_{2}$} when the RF magnetic field strength is sufficiently small to be negligible.

\section{Experimental setup }

The experimental setup is shown in Fig. 1, in which a cubic purified  $\rm{^{87}Rb}$ atomic vapor cell, contains 100 Torr $N_{2}$, with a length of 15 mm was used for the experiment. The AC-driven non-magnetic heating films and Servo loop act as a heating system to control the temperature of the atomic vapor cell at 75 °C. The cell is housed within four layers of $\mu$-metal magnetic shielding to shield the ambient magnetic field. The three-axis Helmholtz coil is placed inside to generate the magnetic field, where the direction of the static magnetic field $B_0$  is along the z-axis, and the static magnetic field $B_0$ is 6.32 $\mu$T.

The magnetometer configuration is shown in Fig. 1(a). The pump light from the 795 nm external cavity semiconductor laser (ECDL) passes through the acousto-optic modulator(AOM), beam expander, Glen-Taylor prism, and quarter-wave plate into a circularly polarized pumped light with a beam diameter of $\sim$ 10 mm, and then enters the rubidium atomic vapor cell along the z-axis. In the experiment, depending on the type of atomic vapor cell, the frequency of the pump light is locked at $\rm{^{87}Rb}$  D1 line F=2-F'=1 transition line. The pump light intensity entering the cell is 12.74 $mW/cm^2$. The probe light from the 780 nm distributed Bragg reflective (DBR) laser also passes through the AOM and the polarizer to become linearly polarized light with a beam diameter of $\sim$ 2 mm. The frequency of the probe light is blue detuned 18 GHz from $\rm{^{87}Rb}$ D2 line F=2-F'=2 transition line, and the light intensity entering the cell is 3.18 $mW/cm^2$. After traversing the cell along the x-axis, the probe light enters a balanced polarimeter where the optical signal is converted into an electrical signal and fed into the LIA. 

For the spin noise spectrum configuration, as shown in Fig. 1(b), the direction of the static magnetic field is along the z-axis. Only a 780 nm linearly polarized probe light is used to traverse the vapor cell along the x-axis. The rotation of the probe light polarization is received by the balanced polarimeter, subsequently into the fast Fourier transform (FFT) to convert the time-domain signal into a frequency-domain signal. 

\begin{figure}[!h]
\centerline{\includegraphics[width=\columnwidth]{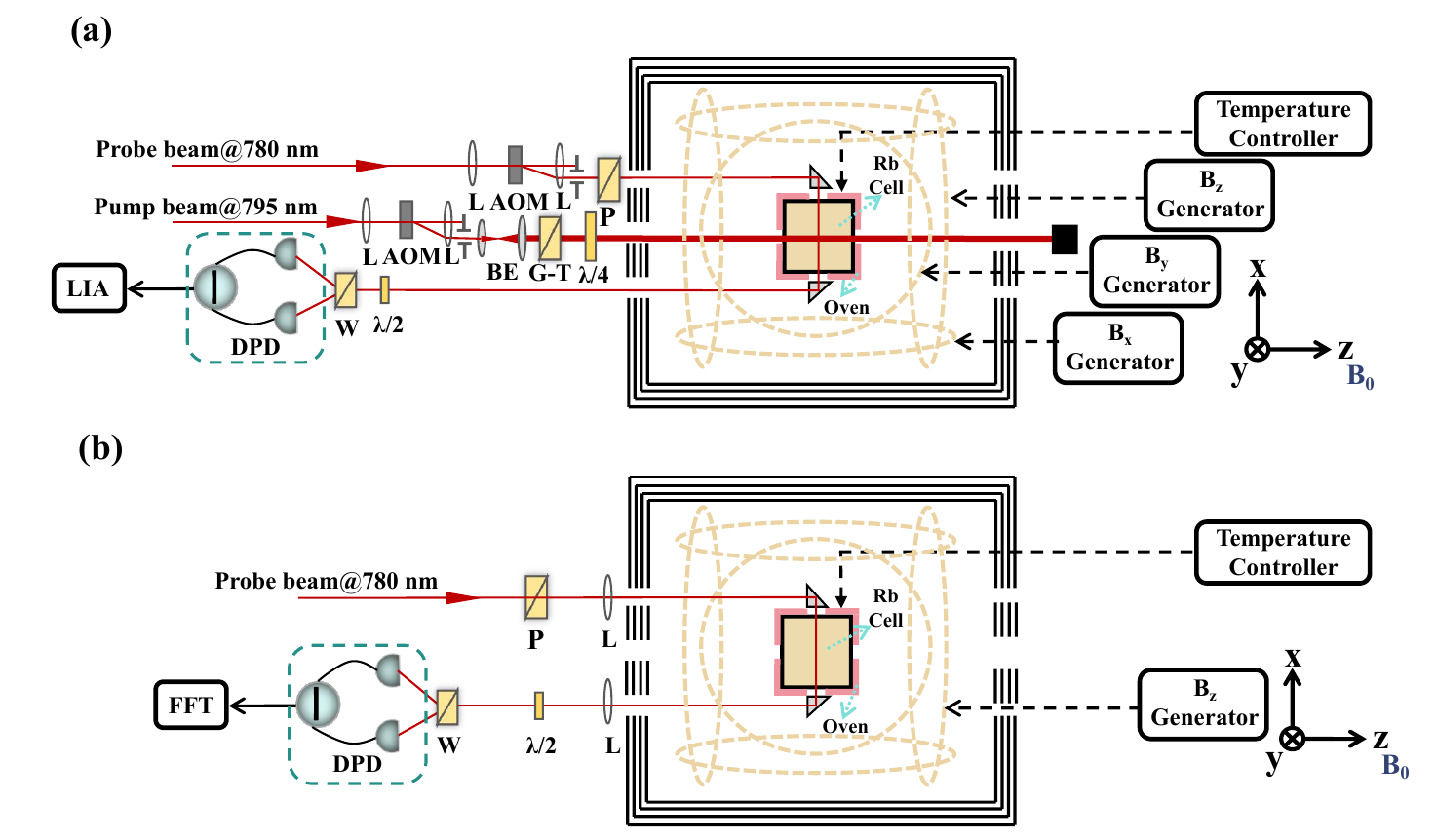}}
\caption{(a) Rubidium atomic magnetometer scheme. (b) Spin noise spectrum scheme. AOM: acoustic-optical modulator; BE: beam expander; $\lambda$/4: quarter-wave plate; $\lambda$/2: half-wave plate; G-T: Glen Taylor Prism; L: lens P: Polarizer; W: Wollaston prism; DPD: Differential photodetector; LIA: Lock-in Amplifier; FFT: Fast Fourier Transform.}
\label{Fig1}
\end{figure}

\section{Results and analysis}

\subsection{Typical four measuring methods results and analysis}
In the spin noise spectrum fitting method, the static magnetic field $B_0$ is 6.32 $\mu$T, and the probe power is 3.18 $mW/cm^2$. After optimizing the parameters, the typical spin noise spectrum signal after FFT is shown in Fig. 2. We set the average sample number to 1000 times, optimized the parameters, and obtained a full width at half maximum(FWHM) of 1.5 kHz by Lorentz fitting.

\begin{figure}[!h]
\centerline{\includegraphics[width=\columnwidth]{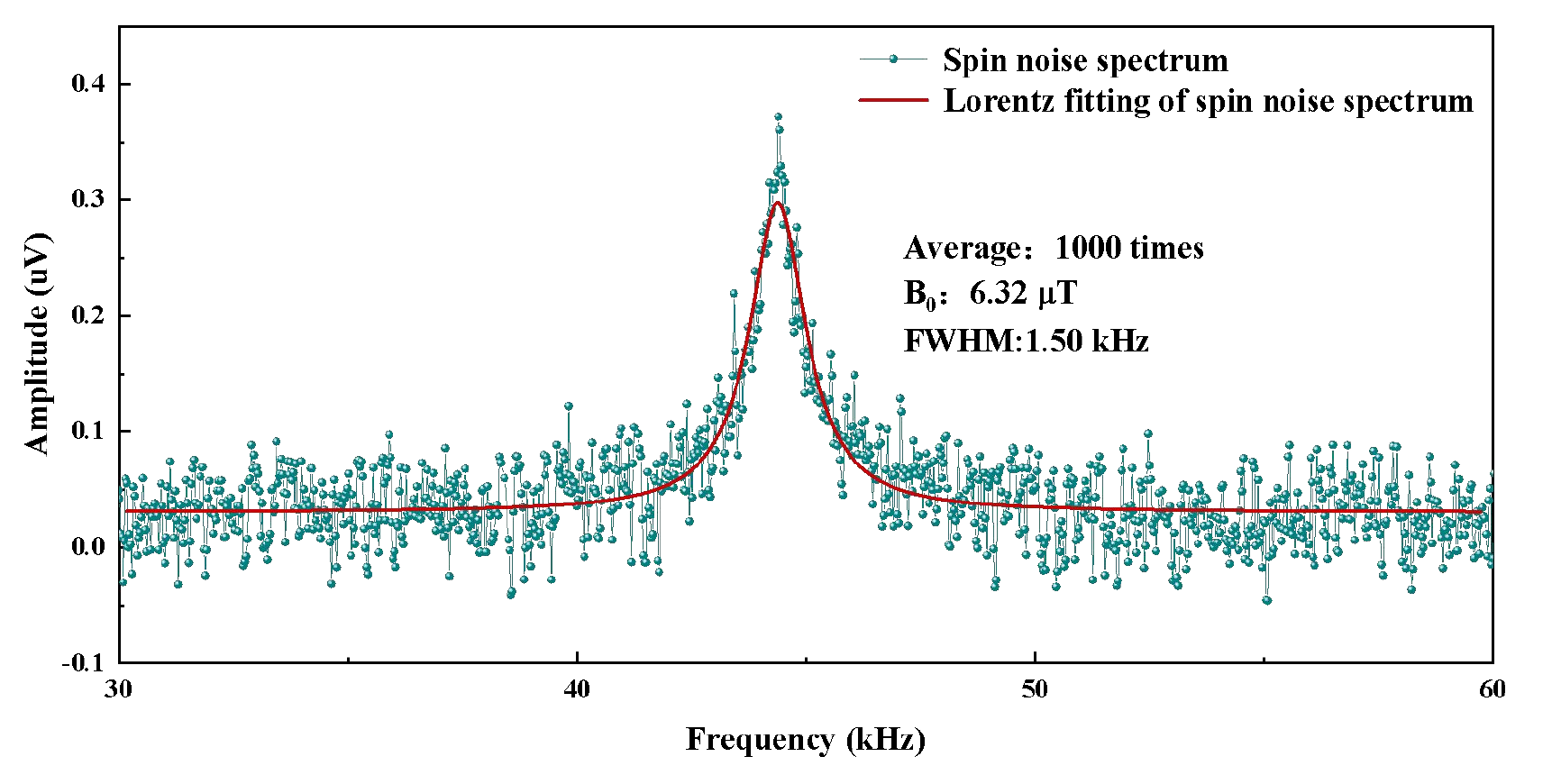}}
\caption{Spin noise spectrum signal. The red curve is the Lorentz fitting, with a typical FWHM of 1.50 kHz. Other parameters: the frequency of the probe beam power is blue detuning 10 GHz from $\rm{^{87}Rb}$ D2 line F=2-F'=2 transition line. The spin noise spectrum is averaged 1000 times and $B_0$ is 6.32 $\mu$T.}
\label{Fig2}
\end{figure}

In the improved FID signal fitting method, we design a timing sequence control system, as shown in the inset of Fig. 3, to separate the pump light, RF magnetic field, and probe light from the time domain, as well as to avoid the influence of crosstalk between the three on the free-induced decay signal[24]. Here, we set the typical timing sequence as 10 ms for the pump light time, 3 mA for the current to apply the RF magnetic field strength, the corresponding $\pi$/2 pulse time of 0.2 ms, and the detection time of 29.8 ms. The typical experimental result is shown in Fig. 3, and the $T_2$ of Rb is 3.7 ms by fitting Eq.(3).

\begin{figure}[!h]
\centerline{\includegraphics[width=\columnwidth]{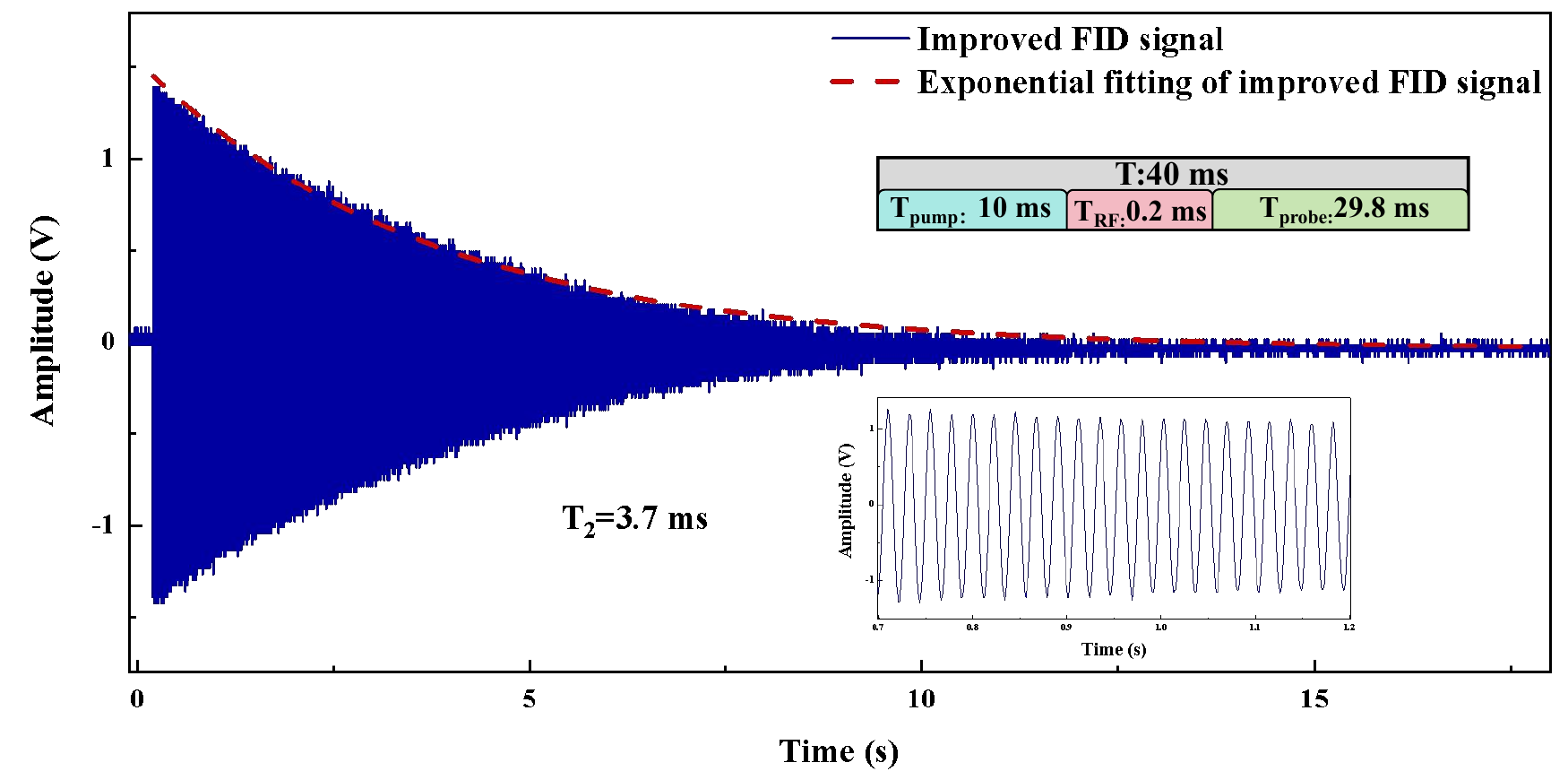}}
\caption{The improved FID signal. The red dotted line is the exponential fitting. Interset: timing sequence control system (top) and several FID precession cycles (bottom).}
\label{Fig3}
\end{figure}

In $\omega_m$ - Broadening fitting method, We set the frequency $\omega_m$ of the modulation field to half the Lamor frequency $\omega_L$ , i.e. n=2, and sweep the frequency near the modulated frequency. Set $B_x$ = 0.41 $\mu$T, $B_y$ = 0 $\mu$T or $B_x$ = 0 $\mu$T, $B_y$ = 0.41 $\mu$T. The typical in-phase and quadrature signals obtained by LIA are shown in Fig. 4, where $\omega_m$ is also the reference frequency of LIA, the extracted $T_2$ of Rb is 3.9 ms.

\begin{figure}[!h]
\centerline{\includegraphics[width=\columnwidth]{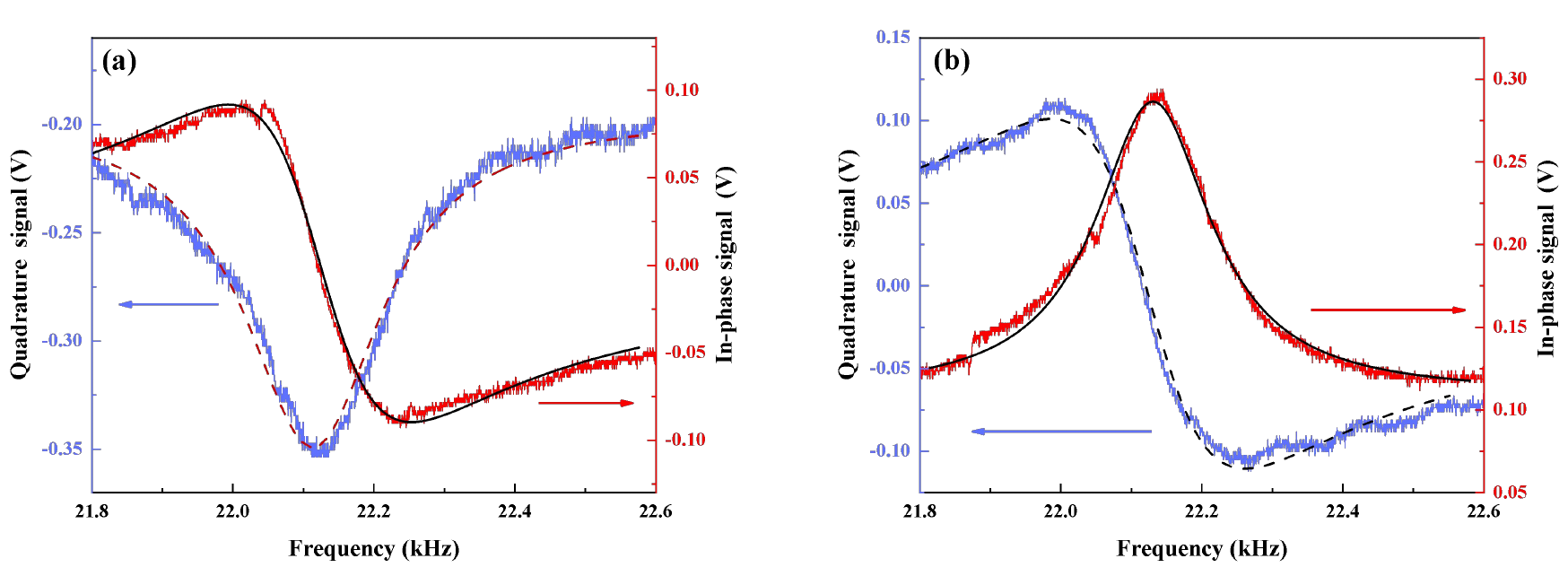}}
\caption{(a) The demodulated in-phase and quadrature signals when $B_x$ = 0.41 $\mu$T, $B_y$ = 0 $\mu$T. (b) The demodulated in-phase and quadrature signals when $B_x$ = 0 $\mu$T, $B_y$ = 0.41 $\mu$T.}
\label{Fig4}
\end{figure}

In magnetic resonance broadening fitting experiments, we apply an RF magnetic field with a frequency equal to the Lamor frequency on the y-axis, and the reference signal frequency demodulated by the LIA is set to the Lamor frequency. By scanning the static magnetic field applied on the z-axis, the in-phase and quadrature signals are shown in Figure 5(a). The FWHM is obtained by Lorentz fitting of the demodulated signal. With the change of RF magnetic field strength, the variation of FWHM with RF magnetic field strength is shown in Fig. 5(b). According to Eq. (6), the $T_2$ of Rb is 3.0 ms

\subsection{Comparison and analysis of different types of rubidium vapor cells}
Furthermore, we selected five types of Rb atomic vapor cells, the parameters of which are shown in Table 1, and used the above four methods to measure the $T_2$ of different Rb atomic vapor cells. Typical results are shown in Fig. 6(a). The figure shows that the 20 mm$\times$20 mm$\times$20 mm purified $\rm{^{87}Rb}$ atomic vapor cell filled with 100 Torr $N_2$ have the best suppression of atomic spin decoherence.

\begin{figure}[!h]
\centerline{\includegraphics[width=\columnwidth]{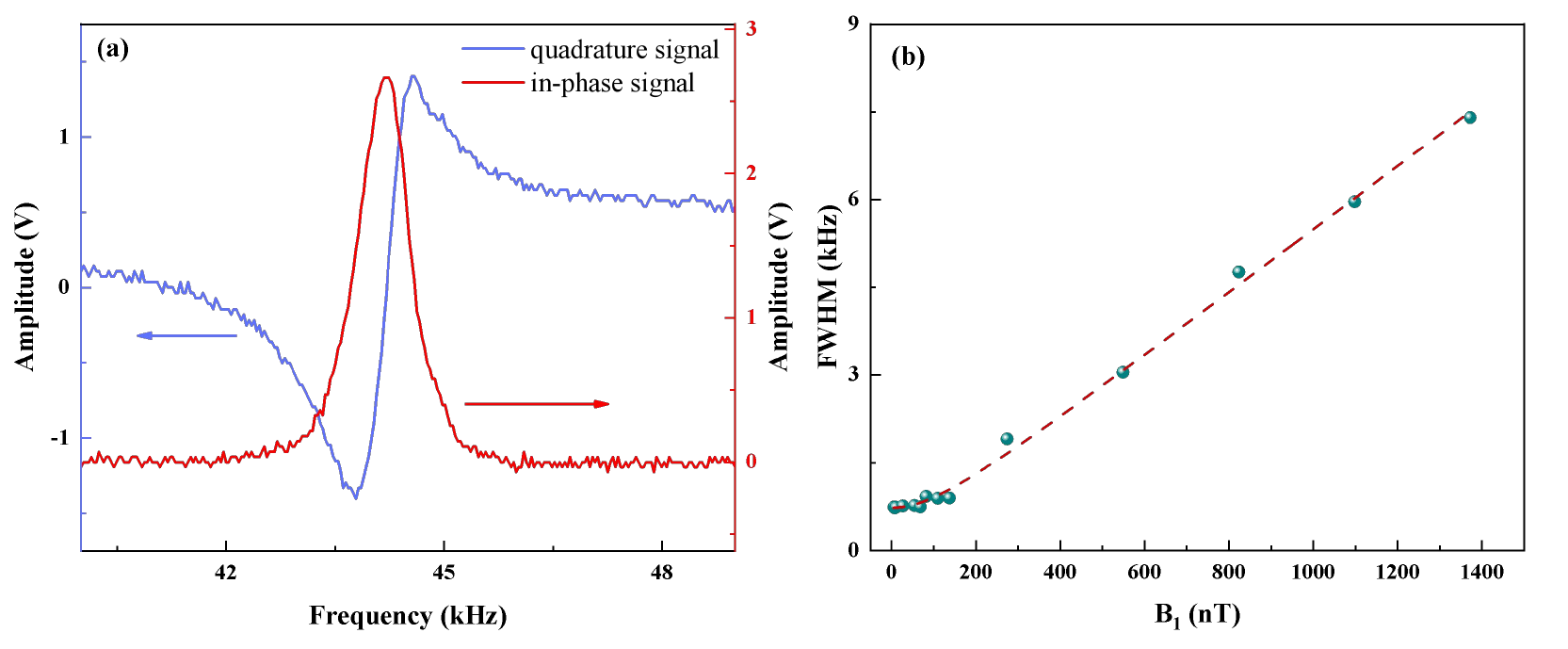}}
\caption{(a) The in-phase and quadrature signals while $B_1$ = 69 nT. (b) Dependence between FWHM and RF magnetic field strength.}
\label{Fig5}
\end{figure}

It is evident from Fig. 6 (a) that the $T_2$ obtained by the spin noise spectrum method is significantly smaller than that of the other three methods. We compare the FWHM obtained by FFT of the improved FID signal fitting method with the FWHM of the spin noise spectrum fitting method, as shown in Fig. 6(b). The FWHM of the spin noise spectrum fitting method is significantly wider than the FWHM of the improved FID signal after FFT. The main reasons are as follows. Firstly, the spin noise spectrum measures atomic spins' fluctuations under unpolarization, which yields a relatively small signal amplitude and a poor signal-to-noise ratio. Simultaneously, the power broadening and the inhomogeneity of the probe light in the Rb cell can have a negative effect. Secondly, due to the influence of the transit time broadening and the measurement time. The longer the measurement time, the more times the atoms cross the probe beam, and the wider the line broadening caused by the crossing time. Furthermore, the inhomogeneity of the static magnetic field also affects the FWHM, further influencing the $T_2$ of Rb.

\begin{table}[!h]
\caption{Specification parameters of the five rubidium atomic vapor cells}
\label{table}
\setlength{\tabcolsep}{3pt}
\begin{tabular}{|p{40pt}|p{55pt}|p{130pt}|}
\hline
Cell number& 
Cell size (mm)& 
Gas filled into cell\\
\hline
NO.1& 
15$\times$15$\times$15& 
$\rm{^{87}Rb}$+100 Torr $N_2$\\
NO.2& 
20$\times$20$\times$20& 
$\rm{^{87}Rb}$+$\rm{^{85}Rb}$+100 Torr $N_2$+20 Torr He\\
NO.3& 
20$\times$20$\times$20& 
$\rm{^{87}Rb}$+100 Torr $N_2$\\
NO.4& 
20$\times$20$\times$20& 
$\rm{^{87}Rb}$+$\rm{^{85}Rb}$+100 Torr $N_2$\\
NO.5& 
20$\times$20$\times$20& 
$\rm{^{87}Rb}$+$\rm{^{85}Rb}$+600 Torr $N_2$+50 Torr He\\
\hline
\end{tabular}
\label{tab1}
\end{table}

In addition to the spin noise spectrum method, the other three methods have differences in measurement results owing to the different magnetic fields at the time of measurement; however, the three methods can be used for different types of magnetometers. Because only a static magnetic field exists in the measurement process and the time-sequence control system avoids the negative effect of pump light in the measurement, the whole measurement process has more cleaner environmentally, and relatively accurate $T_2$ of Rb are obtained from the improved FID signal fitting method.

\begin{figure}[!h]
\centerline{\includegraphics[width=\columnwidth]{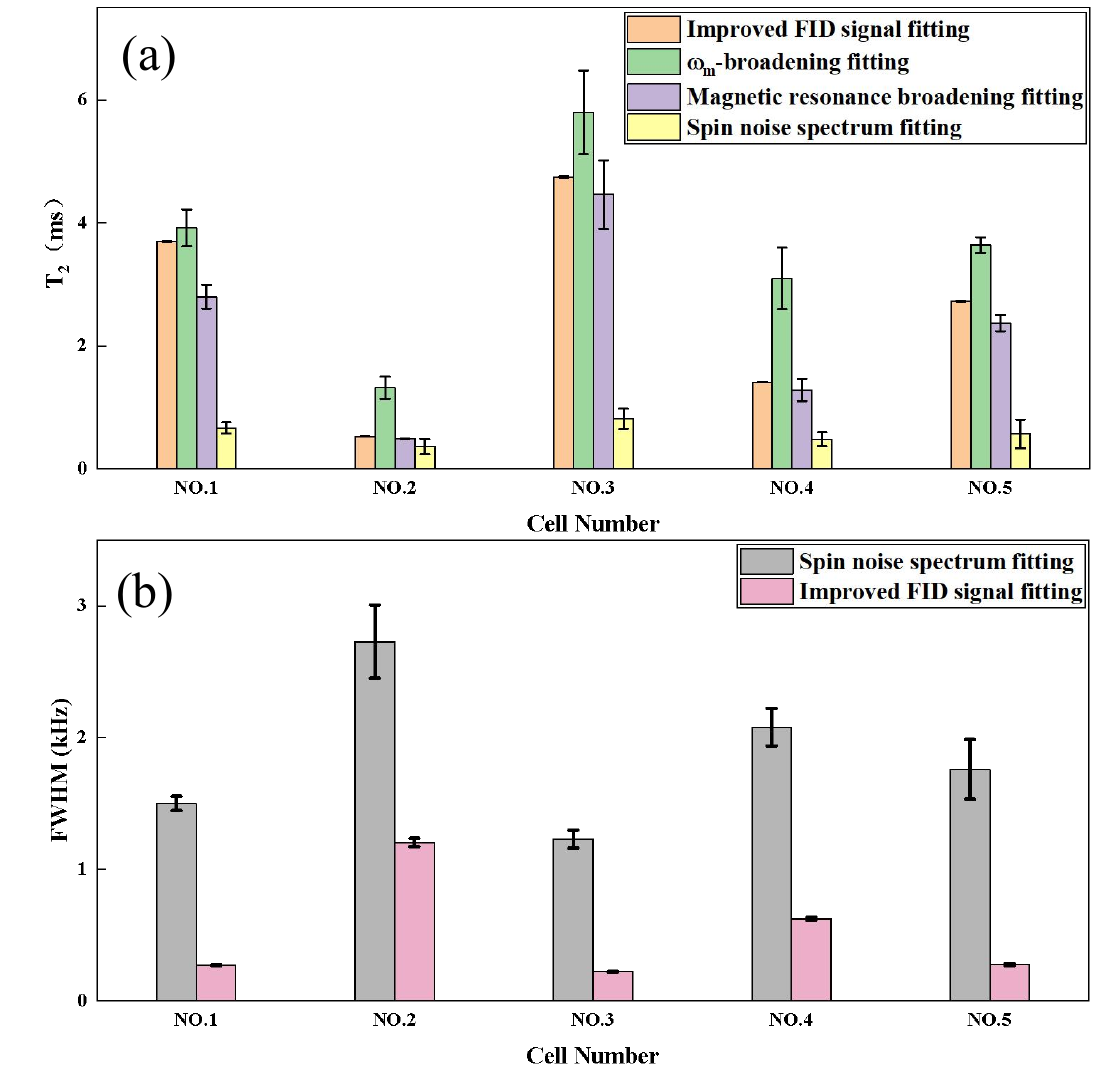}}
\caption{(a) The $T_2$ of Rb measured by improved FID signal fitting, $\omega_m$ - broadening fitting, magnetic resonance broadening fitting, and Spin noise spectrum fitting in different Rb vapor cells. The temperature of Rb vapor is controlled at 75°C. (b) The FWHM of improved FID signal fitting and Spin noise spectrum fitting in different Rb vapor cells. }
\label{Fig6}
\end{figure}

 Based on the improved FID signal fitting method, we experimentally measured the dependence between different temperatures and the $T_2$ of Rb in cell number NO.1, and typical results are shown in Fig. 7. The $T_2$ of Rb is positively related with the atomic number density at low temperatures. However, we filled $N_2$ as a buffer gas and a fluorescence quenching gas to suppress the effect of collisional relaxation(spin-destructive collision and spin-exchange collision) on $T_2$ of Rb. The $T_2$ of Rb gradually decreases with further increasing temperature due to the incompletely eliminated relaxation mechanism, especially the dominant mechanism of spin-exchange collisional relaxation[23].

\begin{figure}[!h]
\centerline{\includegraphics[width=\columnwidth]{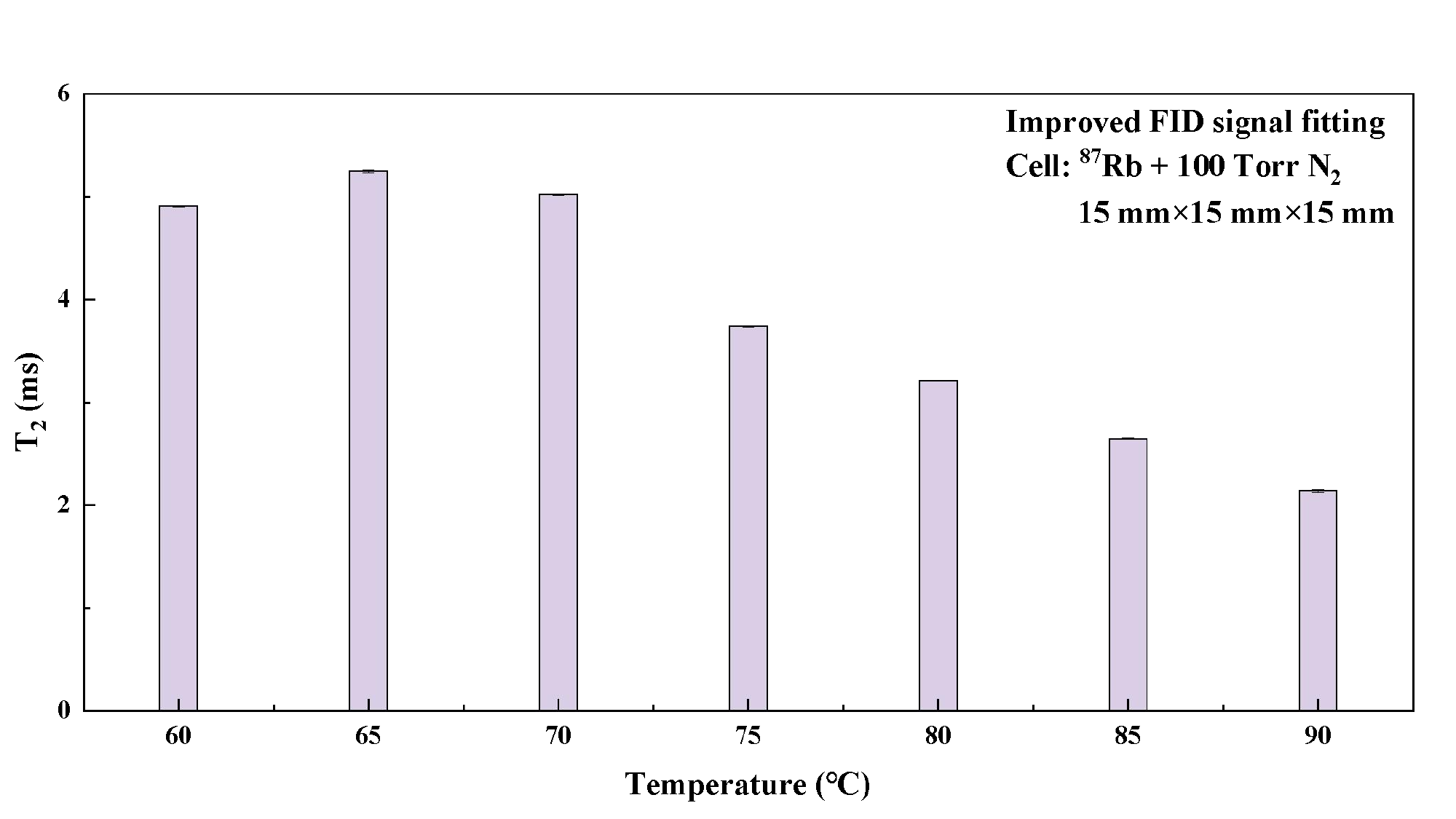}}
\caption{The $T_2$ of rubidium atomic vapor, measured by improved FID signal fitting, varies with the temperature of the rubidium vapor cell. Cell parameters:15 mm$\times$15 mm$\times$15 mm, $\rm{^{87}Rb}$+100 Torr $N_2$}
\label{Fig7}
\end{figure}

\section{Conclusions}
In this paper, four methods were employed to measure the $T_2$ of Rb based on the Rb atomic vapor cell filled with different types and pressures buffer gas. We demonstrate the dependence between different atomic number densities and the $T_2$ of Rb. The effect of the relaxation mechanism on the $T_2$ of Rb was suppressed by filling with buffer gas. 
Spin-exchange collisions can be further suppressed by optical narrowing effects[24],[25],[26] or by heating the cell to extremely high temperatures to keep atoms in the spin-exchange relaxation-free state[27],[28].

Among these four methods, the $T_2$ of Rb obtained by the spin noise spectrum fitting method is less than the actual value owing to the limitation of the measurement technology, and a high-resolution spin noise spectrometer is required for experimental acquisition. However, this method provides a roundabout method for comparing the $T_2$ in different types of atomic vapor cells. The experimental setup for the magnetic resonance broadening fitting method is relatively simple. It is easier to implement experimentally, and even the $T_2$ of atoms can be measured only by measuring the pump light [9]. The  $\omega_m$ - broadening fitting method allows real-time monitoring of the drift of the static magnetic field, which can be compensated by a servo-loop system. Although the experimental setup for the improved FID signal fitting method is relatively complex, it ensures that the negative effects of the RF magnetic field and pump light are not introduced during the measurement by applying a timing sequence control system. In addition, the $T_2$  of atoms measured by this method is relatively accurate.

These four methods can be used in different measuring environments. The $T_2$  of atoms obtained by different measuring methods are slightly different, mainly because the magnetic field environment is different during the measurement.  However, all of them intuitively show the relationship between the buffer gases of different pressures and the $T_2$. This provides a variety of feasible methods for future precision measurement fields, such as the improvement of the sensitivity of atomic magnetometers and the selection of atomic vapor cells. Meanwhile, by comparing the $T_2$ of Rb obtained by five different types of Rb atomic vapor cells under the same conditions, we can obtain an optimal type of atomic vapor cell. This study has important application reference value in the fields of the fabrication of atomic vapor cells, and the Selection of buffer gas types and pressures.

\section*{Funding}

This work was supported by the National Key R \& D Program of China (Grant No. 2021YFA1402002), the National Natural Science Foundation of China (Grant Nos. 11974226), the Shanxi Provincial Graduate Education Innovation Project (Grant No.2022Y022) and the 1331 Key Subjects Construction of Shanxi Province, China.

\section*{Conflict of Interest}

The authors have no conflicts to disclose.

\end{document}